\documentclass[12pt]{article}
\usepackage{epsf}
\usepackage{dcolumn}
\usepackage{bm}
\usepackage{amssymb}
\usepackage{epsfig}
\usepackage{graphicx}
\usepackage{amsfonts}
\usepackage{amssymb}
\usepackage{latexsym}
\def\ee{\end{equation}}
\def\ba{\begin{eqnarray}}
\def\ea{\end{eqnarray}}

\def\bq{\begin{quote}}
\def\eq{\end{quote}}

 at 10truept

\newcommand{\beq}{\begin{equation}}
\newcommand{\eeq}{\end{equation}}
\newcommand{\beqa}{\begin{eqnarray}}
\newcommand{\eeqa}{\end{eqnarray}}
\newcommand{\bea}{\begin{eqnarray}}
\newcommand{\eea}{\end{eqnarray}}

\newcommand{\al}{\alpha}

\newcommand{\mpl}{M_{Pl}}

\newcommand{\overskrift}[1]{\vspace{6.0mm}\noindent\textbf{#1}\vspace{1.5mm}}

\def\lesssim{~\mbox{\raisebox{-.6ex}{$\stackrel{<}{\sim}$}}~}

\def\ltap{\ \raise.3ex\hbox{$<$\kern-.75em\lower1ex\hbox{$\sim$}}\ }
\def\gtap{\ \raise.3ex\hbox{$>$\kern-.75em\lower1ex\hbox{$\sim$}}\ }
\def\gl{\ \raise.5ex\hbox{$>$}\kern-.8em\lower.5ex\hbox{$<$}\ }
\def\roughly#1{\raise.3ex\hbox{$#1$\kern-.75em\lower1ex\hbox{$\sim$}}}

\begin{document}

\thispagestyle{empty}

\begin{flushright}
{\tt HIP-2007-09/TH}\\
{\tt hep-ph/0702236}\\
\end{flushright}

\vskip2cm
\begin{center}
{\Large{\bf The Matrix Reloaded - on the Dark Energy Seesaw}}
\vskip2cm {\large Kari Enqvist${}^1$\footnote{\tt
enqvist@helsinki.fi} Steen Hannestad${}^2$\footnote{\tt
sth@phys.au.dk}  Martin S. Sloth${}^2$\footnote{\tt
sloth@phys.au.dk}}\\

\vspace{.5cm}

\vskip 0.1in

{\em ${}^1$Helsinki Institute of Physics, P.O. Box 64,}\\
{\em FIN-00014 University of Helsinki, Finland}\\

\vskip 0.1in

{\em ${}^2$Department of Physics and Astronomy, University of Aarhus}\\
{\em DK-8000 Aarhus C, Denmark}\\

\vskip 0.1in
\vskip 0.1in
\vskip .25in
{\bf Abstract}
\end{center}
We propose a novel mechanism for dark energy, based on an extended seesaw
for scalar fields, which does not require any new physics at
energies below the TeV scale. A very light quintessence mass is
usually considered to be technically unnatural, unless it is
protected by some symmetry broken at the new very light scale. We
propose that one can use an extended seesaw mechanism to construct
technically natural models for very light fields, protected by SUSY
softly broken above a TeV.
\vfill \setcounter{page}{0} \setcounter{footnote}{0}
\newpage

\setcounter{equation}{0} \setcounter{footnote}{0}

\noindent The evidence for the recent acceleration in the expansion rate of
the Universe is compelling. The effect has been probed directly
using distant type Ia supernovae \cite{perlmutter99,riess98}, the
most recent data coming from the ESSENCE and SuperNova Legacy Survey
(SNLS) projects \cite{astier06,Wood-Vasey:2007jb}. The finding is
confirmed by precision measurements of the Cosmic Microwave
Background (CMB) by the WMAP satellite \cite{Spergel:2006hy}, as
well as measurements of the large scale structure (LSS) from the
Sloan Digital Sky Survey \cite{Tegmark:2006az}.

This host of different observations all point to a cosmology in
which space-time is flat and the energy density dominated by a dark
energy component with an equation of state close to -1. Since
$\rho_{\rm DE} \sim H^2 \sim (10^{-3} \, {\rm eV})^2$ at present,
the dark energy phenomenon is naturally associated with an energy
scale of order $10^{-3}$ eV, much lower than any known scale in
particle physics.

The simplest way of introducing dark energy with the correct
properties is to introduce a very small cosmological constant by
hand, without addressing the origin of this small scale. This is
unsatisfactory, and thus we would like to construct physically well
motivated models where the new small mass scale emerges from a
deeper principle that also leads to testable predictions for
observable quantities such as the equation of state of dark energy.

One such model is the quintessence model
\cite{Wetterich:1987fm,Peebles:1987ek,Ratra:1987rm,Zlatev:1998tr}
where dark energy is driven by the slow-roll of a very light
quintessence field. The effective mass has to be less than the
inverse Hubble scale today, i.e. $m_{eff}\lesssim H^{-1} \sim
10^{-42}$ GeV for this model to work. Such a small mass cannot be
protected from radiative corrections unless some additional symmetry
is invoked. In the Minimal Supersymmetric Standard Model (MSSM),
masses are protected by supersymmetry (SUSY) softly broken at the
TeV scale. This implies that the fields generally gets radiative
mass corrections of the order a TeV.

Compared to this the quintessence mass is incredibly small and even
if added by hand, it will not be stable against radiative
corrections. In neutrino physics a similar problem arises because of
the hierarchy between the known neutrino masses $m_\nu \sim $ eV,
and the TeV scale of the Higgs VEV. However, the seesaw mechanism
\cite{Minkowski:1977sc,Yanagida:1980,Gell-Mann:1980vs,Mohapatra:1979ia,Schechter:1980gr}
 attempts to explain this by introducing a new set of right-handed neutrinos with  Majorana
masses of order $M_{\rm GUT}$. In that case the combined neutrino mass matrix can
be diagonalised to contain two sets of Majorana masses of order $m
\sim v^2/M$ and $M$ for the left- and right-handed components
respectively, where $v \sim$  1 TeV. This yields $m_{\nu_L} \sim$ eV,
the smallest non-vanishing mass we would naively expect to be radiative stable,
when couplings to known particles in the standard model are taken
into account.

In the case of scalar fields one can introduce a new shift symmetry which requires the mass to
vanish identically. By breaking the shift symmetry explicitly by a small
parameter, $\mu$, one can obtain a radiatively stable small effective
mass. To obtain dark energy of the right magnitude, one needs
$\mu\approx 10^{-3}$eV. In the pseudo Nambu-Goldstone Boson  (pNGB)
model \cite{Frieman:1995pm} there is no explanation of where this
scale is emerging from. In the case of the Peccei-Quinn axion,  an
effective $\mu\approx \Lambda_{QCD}$ is generated by QCD instanton
effects, and in order
to explain the origin of the low symmetry breaking scale $\mu\approx
10^{-3}$eV, in ref.~\cite{Hall:2005xb} it was proposed that the pNGB quintessence mass
$\mu$ could have a similar dynamical  origin.  However, the protection of such a small scale would
require symmetry breaking that takes place in a hidden sector.

Here we will consider an alternative possibility. The light neutrino
mass eigenvalue in the usual neutrino seesaw mechanism is of order
$v^2/M$. However, as we show below, there exist extended mass
matrices of higher dimensions where the smallest mass eigenvalues
are $v(v/M)^n$ with $n\geq 0$ is an integer.  Moreover, we present explicit mass
matrices of higher dimensions where all the entries in the diagonal
are non-zero and of order $M^2$, but which nevertheless yield eigenvalues of
the type $v(v/M)^n$. This has applications to scalar field models,
where the diagonal elements are naturally of order $M$ and the
off-diagonal entries can be generated by soft SUSY breaking terms of
order $v$ or larger. Requiring that the resulting light mass is the
quintessence mass implies that, in the case of $n=5$, the heavy mass
scale of the scalars coincides roughly with the required mass scale
in the simplest scenario of chaotic inflation, $M \sim 10^{13}$ GeV.
This adds the intriguing possibility that the heavy fields can also
have been responsible for inflation in the early universe.

In the present paper we present one particular example of a
configuration of 8 complex scalar fields which by virtue of the
extended seesaw mechanism has a composite flat direction with
properties adequate for quintessence. We also discuss the possible
physical origin of such a mass matrix and construct one specific
example, where the seesaw scalar mass matrix arises from a
particular configuration of 5-, 6-, and 7-branes embedded in a 10D
space-time.

\bigskip

The conventional two-neutrino seesaw model involves a mass matrix with one
vanishing element in the diagonal. This is naturally achieved for
fermions since chiral symmetry prevents them from acquiring a
mass. However, the masses of scalar fields are not in general
protected by a similar symmetry, and we expect scalar fields to
acquire masses at least of order $v^2$.  If we desire to construct a
seesaw mechanism for scalar fields, this implies that we need to go
to mass matrices of higher dimension.

First we observe that with a sufficiently large mass matrix with
entries $m$, $M$ where $m<<M$, we can possibly have a mass eigenvalue
$m^5/M^4$. In this case we can take $m\approx v$ and $M= M_{GUT}$ in
order to obtain the desired quintessence mass scale. In this case a
theory of the form
 \beq \label{pot1}
 V(\phi) = \sum_{ij}m^2_{ij}\phi_i\phi_j~,
 \eeq
 with all entries $m_{ij} \gtrsim v$
protected by supersymmetry, would suffice. Naively, such a mass
matrix would give us a lightest mass of the expected size $m^2/M$,
but for certain configurations the lightest mass eigenvalue is $m^5/M^4$,
which would then be a radiatively stable quintessence mass.

Naively, this is not easy to realise in a realistic SUSY model. If
the mass of the effective quintessence field is of the order of the
inverse Hubble scale today $m_q\approx H_0$, then the vev of the
quintessence field must be Planck scale $\langle q \rangle \approx
\mpl$. This implies that we also need to  assume that self-couplings
induced by marginal and irrelevant operators are absent in order to
obtain a right energy density.

The next question is thus how to generate mass terms and mixings without
generating marginal self-coupling operators.  Consider for simplicity a
superpotential of chiral superfields, which are singlets under the
SM charges, given by
 \beq
 W = \sum M \Phi_i\Phi_j~.
 \eeq
This induces a scalar potential for the scalar components $\phi_i$
of the superfields $\Phi_i$ with
supersymmetric mass terms of the type $|dW/d\phi|^2$. In addition,
there are
soft SUSY breaking terms, which we assume to have the form
${(m^2)^{i}}_j\phi_i\phi^{*j}$, thus effectively neglecting the
contribution from the A-term $\sim {\rm Re}(A_{ij}\phi_i\phi_j)$ for simplicity
(recall that in e.g. no-scale supergravity $A_{ij}=0$). Thus, the scalar potential would
have the generic form
 \beq
 V = \sum_{ij} \left[ M^2|\phi_i|^2+M^2
|\phi_{j}|^2+{(m^2)^{i}}_j\phi_i\phi^{*j} \right]
 \eeq

In this model there are $N$ complex scalar fields of mass $M$. In
order to generate a configuration with an $m^5/M^4$ flat direction,
we need $N\geq7$. However, in order to sufficiently decouple the
flat direction, it is necessary to go to $N=8$, as we will see
below.

Consider a model with 8 complex fields fields, which have masses set
by some heavy scale near the GUT scale $M\approx M_{GUT}$. Then, if we assume that a symmetry
breaking mechanism at the GUT scale induce bilinear mixings for some
of the scalars at scale $M$ and also that soft SUSY breaking bilinear
mixing terms are induced at the weak scale $m\approx v$, we can obtain a mass
matrix of the form
 \beq
 \mathcal{M}^2 =
\left(\begin{array}{cccccccc}
M^2 & m^2 & 0 & M^2 & 0& m^2 & m^2 & m^2 \\  m^2 & M^2 & 0 & m^2 & m^2 & m^2 & 0 & 0 \\  0 & 0 & M^2 & 0 & M^2 & m^2 & 0 & 0 \\
M^2 & m^2 & 0 & M^2 & M^2 & M^2 & 0 & 0 \\  0 & m^2 & M^2 & M^2 & M^2 & M^2 & 0 & m^2 \\ m^2 & m^2 & m^2 & M^2 &  M^2 & M^2 & 0 & m^2 \\
m^2  & 0 & 0 & 0 & 0 & 0 & M^2 & M^2 \\
m^2 &  0 & 0 & 0 & m^2 & m^2 & M^2 & M^2
 \end{array}\right)~.
 \label{massmatrix}
  \eeq
Some of the off-diagonal elements have here been chosen to vanish.
This is necessary in order to obtain a mass eigenvalue of the order
$m_{flat}\approx m^5/M^4$ and thus an effectively light scalar that
can act as quintessence. We have arrived at the form Eq.
(\ref{massmatrix}) by numerical inspection, and will discuss the
physical origin of the zeros shortly. However, let us first note
that the mass matrix Eq. (\ref{massmatrix}) may be diagonalised by
the matrix $P$, such that $\mathcal{M}_d= P^{-1}\mathcal{M}P$. For
Eq. (\ref{massmatrix}) we find numerically the eigenvalues as
 \beq
 \mathcal{M}^2_d =  diag.(a_1 M^2,  a_2 M^2, a_3 M^2, a_4 M^2, a_5 M^2, -a_6 M^2,
  -a_7 M^2,  a_8  (m^5/M^4)^2)~,
  \label{massdiagonal}
  \eeq
where the
numerical coefficients $a_1,\dots,a_8$ are positive and of order
one.

As can be seen from Eq. (\ref{massdiagonal}), there are two negative
eigenvalues which give rise to tachyonic instabilities unless we
stabilise them with further interactions. We could try to add
Yukawa's in the tachyonic directions, but this will in general shift
the vacuum of the fields away from zero and lift the mass of the
light field.  So we need to be more careful. If the interaction
eigenstates are $\phi_1,\dots,\phi_8$ and the mass eigenstates are
$\chi_1,\dots,\chi_8$, then the mass squared eigenvalue of the mass
eigenvector $\chi_8$ is  order $(m^5/M^4)^2$. We can write
 \beq
\chi_8 = \sum_{i=1}^{8}r_i \phi_i~.
 \eeq
In the example of the
$8\times 8$ matrix above, it turns out that $r_2 \approx
(m^7/M^7)^2$, which implies that the light direction has a very
suppressed contribution from $\phi_2$. Using the matrix $P^{-1}$ we
can invert this linear system and find $\phi_2 = t_1\chi_1 +\dots +
t_8\chi_n$ with $t_8\approx r_2 \approx  (m^7/M^7)^2$. Thus, if we
lift the potential by a term $\lambda M^4(\phi_2/M)^{n}$, it will remain
flat in the $\chi_8$ direction.

Thus we are led to adding to the potential a non-renormalisable
contribution
 \beq \lambda
M^4\left(\frac{\phi_2}{M}\right)^{n}=\lambda
M^4\left(t_1\frac{\chi_1}{M}+\dots
+t_{8}\frac{\chi_{8}}{M}\right)^{n}~.
 \eeq
Now, due to the negative mass eigenstates, we assume that all fields
acquire a true vacuum expectation value of order $M$. The vacuum
energy in the new true vacuum is order $\lambda^{-1/(n-2)}M^4$. We
assume that by renormalisation we can subtract the true vaccum
energy of the theory. The only non-zero contribution left is from
the flat direction, which is displaced from its true vacuum. The
dominant contribution to the potential of the flat $\chi_8$
direction is given by
 \beq V(\chi_8) \approx a_8
\left(\frac{m^5}{M^4}\right)^2\chi_8^2 +\sum_{i=1}^7\lambda M^4
t_i^{n-1}t_8 \frac{\chi_8}{M}\left(\frac{\chi_i}{M}\right)^{n-1}~.
 \eeq
Thus, we require $t_8 \ll (m^5/M^5)^2$ in order to suppress the
contribution from the Yukawas.  In the example above we found $t_8
\sim \mathcal{O}((m^7/M^7)^2)$ and the dominant contribution to the
$\chi_8$ potential is from the very light mass term -- light enough
to sustain quintessence.

We have found that in models with 7 scalar fields it is impossible to achieve mixings
smaller than $(m^3/M^3)^2$ which is insufficient to maintain the
lightness of the flat direction. Therefore a model with 8 complex
scalar fields appears to be the simplest possible configuration in
which the smallness of the quintessence mass can be protected
against the Yukawa coupling used to cure the tachyonic instability.
In models with 8 fields, our particular configuration Eq. (\ref{massmatrix}) is
not unique but there are also other possibilities for having $m_{\rm flat}
\sim m^5/M^4$ and $r \sim (m^7/M^7)^2$.

Given the extreme smallness of the mass along the flat direction and
its very small mixing one might worry about the stability against
small corrections to individual elements in the matrix. In order to
understand the stability of the small eigenvalues and mixings, it is
useful to note that when the soft mass is taken to zero, $m\to 0$,
three discrete $\mathbb{Z}_2$ symmetries are restored:
 \bea
\textrm{Sym. 1}&:& ~\phi_2 \to -\phi_2\nonumber\\
\textrm{Sym. 2}&:& ~\phi_7 \to -\phi_7~,~~\phi_8 \to -\phi_8\nonumber\\
\textrm{Sym. 3}&:& ~ \phi_1 \to -\phi_1~, ~~\phi_3 \to -\phi_3~, ~~\phi_4 \to -\phi_4~, ~~\phi_5 \to -\phi_5~,~~ \phi_6 \to -\phi_6~,
 \eea
This implies that all the $M$ values
can be perturbed by a factor ${\mathcal O}(1)$, as long as they are
identical within each of the sub-blocks defined by the symmetries.
Furthermore, all the $m$ values along rows or columns within a
sub-block can be perturbed by a factor of order unity without
destabilising the hierarchy. For example, all the $M$ elements along
rows or columns $13456$ can be changed by a factor of order unity
without changing the structure of the eigensystem.

Since a non-zero value for $m$ breaks all discrete
symmetries of the mass matrix, we may worry wether it is possible to
prevent the zero entries in the mass matrix from obtaining values of
order $m$. It is beyond the scope of this paper to construct a fundamental model
that gives rise to the mass matrix Eq. (\ref{massmatrix}), but as an example of how a mass matrix with the desired
properties might emerge in an effective theory, let us consider the following scenario
that makes use of branes in extra dimensions.

Suppose that each of the complex scalar fields are each
quasi-localised on each their respective branes, with wave functions
in the extra dimensions proportional to $\exp(-|y-y_i|M)$,  where
$y_i$ denotes the location of the brane.  If the complex scalars
carry the same charge, $Q$, under a global $U(1)$ symmetry  $\phi_i
\to \exp(iQ\al)\phi_i$, then when all the branes are on top of each
other, this allows for bilinear mixing terms of the type
$(m^2)^i_j\phi_i\phi^{*j}$. However, if the branes are
geographically separated in the extra dimensions, the overlap of the
corresponding scalar fields' wavefunctions is exponentially
suppressed. This implies that, when the extra dimension is
integrated out, the bilinear interactions of the effective theory
are exponentially suppressed and became of the form
$(m^2)^i_j\exp(-Mr)\phi_i\phi^{*j}$, where $r$ denotes the brane
separation distance. From the point of view of the four dimensional
low energy effective theory, when the branes are separated it
appears as if there are eight distinct $U(1)$ symmetries, and the
flavor violating interactions that breaks the $U(1)\otimes
U(1)\otimes\dots\otimes U(1)$ are exponentially suppressed, while
there is no exponential suppression if the corresponding branes are
on top of each other \cite{Dvali:1999gf}.

If we assume that all the elements $(m^2)^i_j$ are given by the
scale $M$, the suppression of the bilinear interaction terms is
given by $M^2\exp(-Mr)$. With $M\approx M_{GUT}$, the branes can be
taken to lie on top of each other in the directions where bilinear
elements in the mass matrix of scale $M$ are induced, while the
branes are separated by $r\approx 60/M$ in the directions where
bilinear elements at the soft scale $m\approx v$ are induced. In the
directions where there are zeros in the off-diagonal, the branes are
separated by a distance $r \gg 262/M$. If we assume that there are
three brane fixed points, $A$, $B$ and $C$, in each dimension, with
$A$ and $B$ separated by $r\approx 60/M$ and $C$ further away
separated at a distance $r \gg 262/M$, this leads to the brane
configuration for eight branes in six extra dimensions shown in
table 1. For instance, in dimension one we will have Brane 1 and
Brane 4 localised on top of each other at fix point $A$ while Brane
2, 7 and 8 are localised on top of each other at the fixed point
$B$. Brane 3 is localised at the fixed point $C$ in this direction.
The remaining branes are not localised in this direction.

\begin{table}[htdp]
\begin{center}\begin{tabular}{| c||c|c|c|c|c|c|c|c|}  \hline& Brane 1 & Brane 2 & Brane 3 & Brane 4 & Brane 5 & Brane 6 & Brane 7 & Brane 8 \\\hline \hline Dim. 1 & A & B &  C & A &  &  & B & B \\\hline Dim. 2 & A &  &  &  & C &  & A &  \\\hline Dim. 3 & A & A & A &  &  & B &  & A \\\hline Dim. 4 &  & A &  &  & B & B & C &  \\\hline Dim. 5 &  &  &  &  & A & A &  & B \\\hline Dim. 6 &  &  A & A & A &  &  & C & C  \\ \hline \end{tabular} \caption{Brane configuration}
\end{center}
\label{defaulttable}
\end{table}

\bigskip

To summarise, we have constructed an extended quintessence mass
seesaw (EQMS) mechanism for a radiatively stable quintessence mass.
A novelty in this model is that it only involves new physics at the
TeV scale and the GUT scale, and the light quintessence mass is
protected by SUSY softly broken above the TeV scale.

We also note that our configuration holds the potential for
providing inflation during SUSY breaking at the $M$ scale. We have  available 8
 potentials of the type $M^2 \phi^2$, for which $M \sim
10^{13}-10^{14}$ GeV, exactly the range needed for $m^2 \phi^2$
chaotic inflation \cite{Linde:1983gd,Lyth:1998xn}.  The requirement
is just that the true vacuum of the system is shifted to $\langle
\phi \rangle = M$ instead of 0, so that the fields slowly roll
towards this value.

As an interesting curiosity, one may note that an eight-by-eight
matrix is the smallest which can provide the particular features
that we require of the quintessence seesaw mass matrix, yielding
small enough eigenvalues and mixings. However, to realise such a
matrix with the type of brane configurations discussed in the
previous section, we must require exactly six extra dimensions,
which is also the generic prediction of string theory.

\overskrift{Acknowledgments}

\noindent KE is partially supported by the Ehrnrooth Foundation and the
Academy of Finland grant 114419



\end{document}